# Latest results from the NA57 experiment

G. E. Bruno on behalf of the NA57 Collaboration [1]

*Dipartimento IA di Fisica dell'Università e del Politecnico di Bari and INFN, Bari, Italy*

**Abstract.** The NA57 experiment at the CERN SPS has measured the production of strange and multi-strange particles in Pb-Pb (and p-Be) collisions at mid-rapidity. The collective dynamics of the collision is studied in the transverse and longitudinal directions as a function of centrality and beam momentum and analysed based on hydrodynamical models. Central-to-peripheral nuclear modification factors at 158 $A$ GeV/$c$ are presented and compared with other measurements and theory.

**Keywords:** QGP, Heavy Ions, Strange particles, collective dynamics
**PACS:** 12.38.Mh, 25.75.Nq, 25.75.Dw

## INTRODUCTION

The measurement of strange particle production provides a powerful tool to study the dynamics of the reaction in heavy-ion collisions. In particular, an enhanced production of strange particles in nucleus–nucleus collisions with respect to proton–induced reactions was suggested long ago as a possible signature of the phase transition from colour confined hadronic matter to a Quark Gluon Plasma (QGP) [1]. Results on hyperon enhancements at 160 and 40 $A$ GeV/$c$ obtained by NA57 can be found in reference [2] and they will not be discussed in this paper.

Important insights into the reaction dynamics can been determined from the $p_T$ distributions of strange particles: at low transverse momenta ($p_T \lesssim 2$ GeV/$c$) they allow to study the transverse expansion of the collision [3]; at higher values ($p_T \gtrsim 2$ GeV/$c$) the $p_T$ dependence of the nuclear modification factors can be used to probe the properties of the medium produced after the collisions [4, 5].

Rapidity distributions provide a tool to study the longitudinal dynamics. Hydrodynamical properties of the expanding matter created in heavy ion reactions have been discussed by Landau [6] and Bjorken [7] using different initial conditions. In both scenarios, thermal equilibrium is quickly achieved and the subsequent isentropic expansion is governed by hydrodynamics.

## COLLECTIVE DYNAMICS

The presence of strong collective dynamics in relativistic heavy-ion collisions has been widely proven and its features measured (see, e.g., ref. [8, 9] for reviews). While longitudinal 'flow' is not necessarily a signature for nuclear collectivity, because it may be due simply to incomplete stopping of the two colliding nuclei, transverse collective expansion flow can only be driven by the transverse pressure gradients between the hot center of the fireball and the surrounding vacuum. The relative extent of the longitudinal flow with respect to the transverse flow can be used to determine the amount of nuclear stopping.

---

[1] For the author list see http://wa97.web.cern.ch/WA97/NA57authors/index.html.

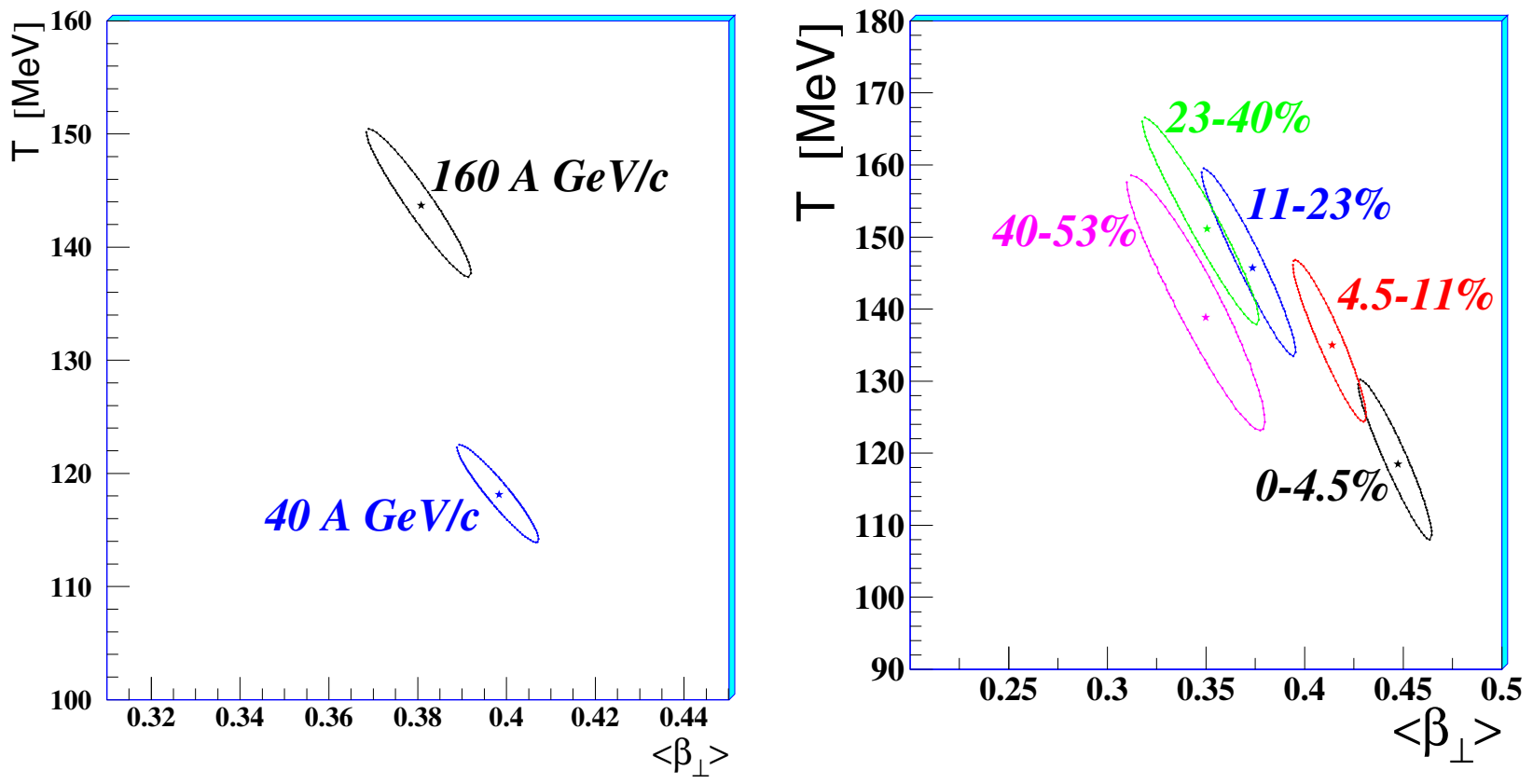

**FIGURE 1.** Contour plots in the $\langle\beta_\perp\rangle$–$T$ plane at the $1\sigma$ condifence level. Left: energy dependence for the most central 53% of the Pb-Pb interactions. Right: centrality dependence at 160 $A$ GeV/$c$.

## Transverse dynamics

The *transverse* dynamics of the collisions has been studied at 160 and 40 $A$ GeV/$c$ respectively in ref. [10] and [11] from the analysis of the transverse momentum distributions of the strange particles in the framework of the blast-wave model [3]. Here we recall some results of this analysis. In the model the double differential cross-section for a particle species $j$ assumes the form:

$$\frac{d^2N_j}{m_\mathrm{T}dm_\mathrm{T}dy} = \mathscr{A}_j \int_0^{R_G} m_\mathrm{T} K_1\left(\frac{m_\mathrm{T}\cosh\rho}{T}\right) I_0\left(\frac{p_\mathrm{T}\sinh\rho}{T}\right) r\,dr \qquad (1)$$

where $m_\mathrm{T} = \sqrt{p_T^2+m^2}$, $\rho(r) = \tanh^{-1}\beta_\perp(r)$, $K_1$ and $I_0$ are two modified Bessel functions and $\mathscr{A}_j$ is a normalization constant. The parameters of the model are the thermal freeze-out temperature $T$ and the transverse flow velocity $\beta_\perp$, whose average has been computed assuming a linear transverse profile $\beta_\perp(r) = \beta_S\left(\frac{r}{R_G}\right)$ [10].

In fig. 1 we show the energy (left panel) and the centrality (right panel) dependences of the freeze-out parameters obtained from the combined fits of the strange particles $m_\mathrm{T}$ spectra to formula 1. With increasing centrality, the transverse flow velocity increases and the freeze-out temperature decreases; the temperature also decreases at lower energy.

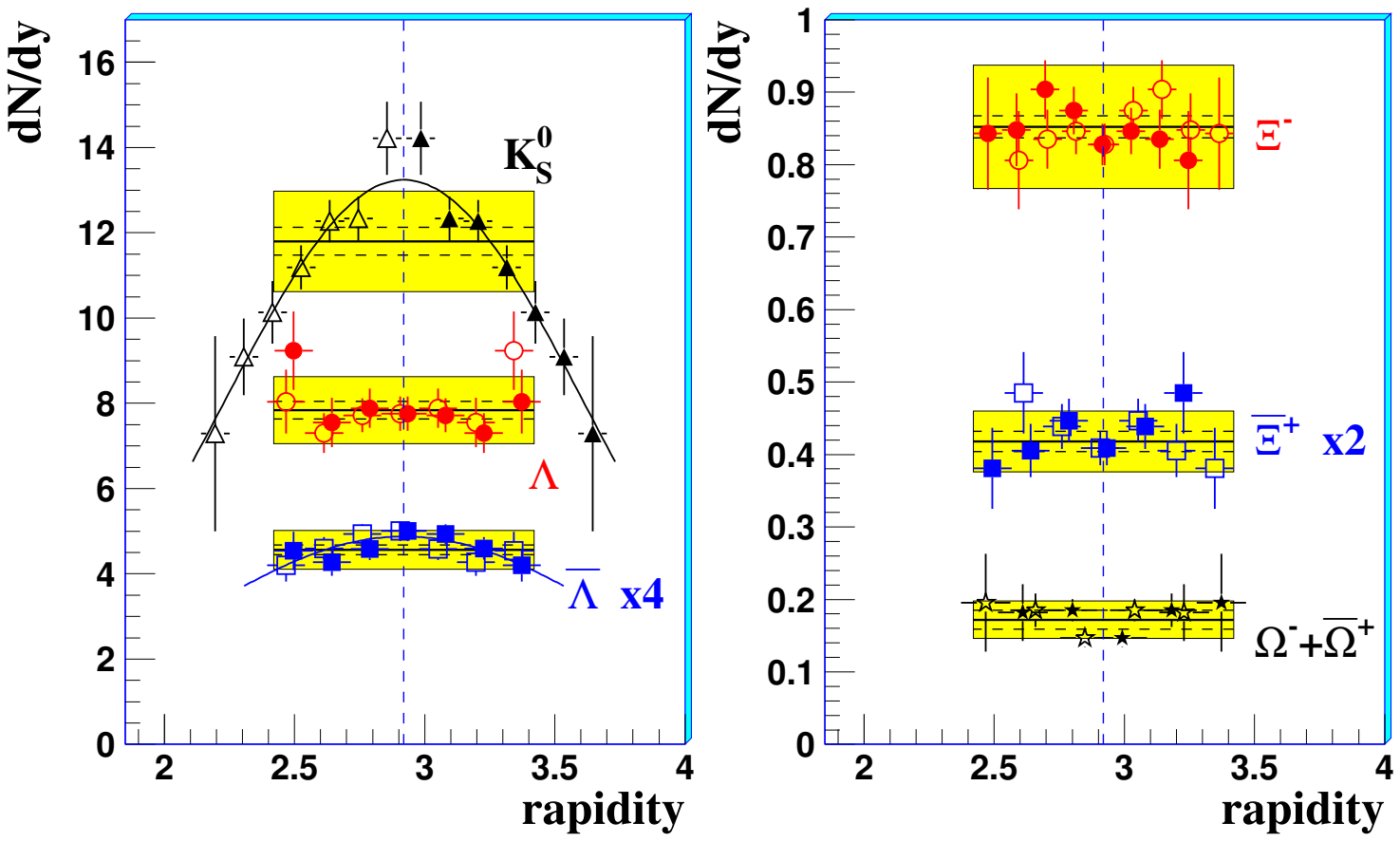

**FIGURE 2.** (colour online) Rapidity distributions of strange particles in the most central 53% of Pb-Pb interactions at 158 *A* GeV/*c*. Closed symbols are measured data, open symbols are measured points reflected around mid-rapidity. The $\overline{\Lambda}$ and $\overline{\Xi}^+$ results have been scaled by factors 4 and 2, respectively, for display purposes. The superimposed boxes show the yields measured in one unit of rapidity (as published in ref. [2]) with the dashed and full lines indicating the statistical and systematic errors, respectively.

## Longitudinal dynamics

Results on the rapidity distributions of strange particles are discussed in detail in ref. [12]; here we outline the main results of the analysis.

The measured rapidity distributions for the centrality range corresponding to the most central 53% of the Pb-Pb inelastic cross-section (total sample) are shown in fig. 2 with closed symbols. The symmetry of the Pb-Pb colliding system allows us to reflect the rapidity distributions around mid-rapidity (open symbols in fig. 2). The rapidity distributions of $\Lambda$, $\Xi^-$, $\overline{\Xi}^+$and $\Omega^-$+$\overline{\Omega}^+$ are compatible, within the error bars, with being flat within the NA57 acceptance window. For the $K^0_S$ and $\overline{\Lambda}$ spectra, instead, we observe a rapidity dependence. The rapidity distributions for these particles are well described by Gaussians centered at mid-rapidity.

The rapidity distributions can be used to extract information about the *longitudinal* dynamics. We use an approach outlined in ref. [3] (i.e., the same blast-wave model used for the study of the transverse dynamics) and [13], where, respectively, Bjorken [7] and Landau [6] hydrodynamics are folded with a thermal distribution of the fluid elements.

In fig. 3 (left panel) the measured rapidity distributions are compared with the expectation for a stationary thermal source and with a longitudinally boost-invariant superposition of multiple isotropic, locally-thermalized sources (i.e. Bjorken hydrodynamics). The average longitudinal flow velocity is evaluated from the fit to be $<\beta_L>=0.42\pm0.03$ with $\chi^2/ndf = 28.2/32$. The freeze-out temperature has been fixed to the value $T_f = 144$ MeV obtained from the analysis of the transverse expansion. The average *transverse* flow velocity has been determined to be $<\beta_\perp>=0.38\pm0.02$, i.e. slightly less than the *longitudinal* flow velocity obtained from the same data sample; this indicates substantial stopping of the incoming nuclei.

In the Landau model the width of the rapidity distribution is sensitive to the speed of sound and to the ratio of the freeze-out temparature to the initial temperature. Landau hydrodynamics can also reproduce simultaneously the distributions for all the strange particles considered ($\chi^2/ndf \simeq 28/32$) but we are not able to put stringent constraints on both the speed of sound and the ratio $T_f/T_0$. The confidence level contours in the $c_s^2$ vs. $\frac{T_f}{T_0}$ parameter space are shown in the right-hand panel of fig. 3.

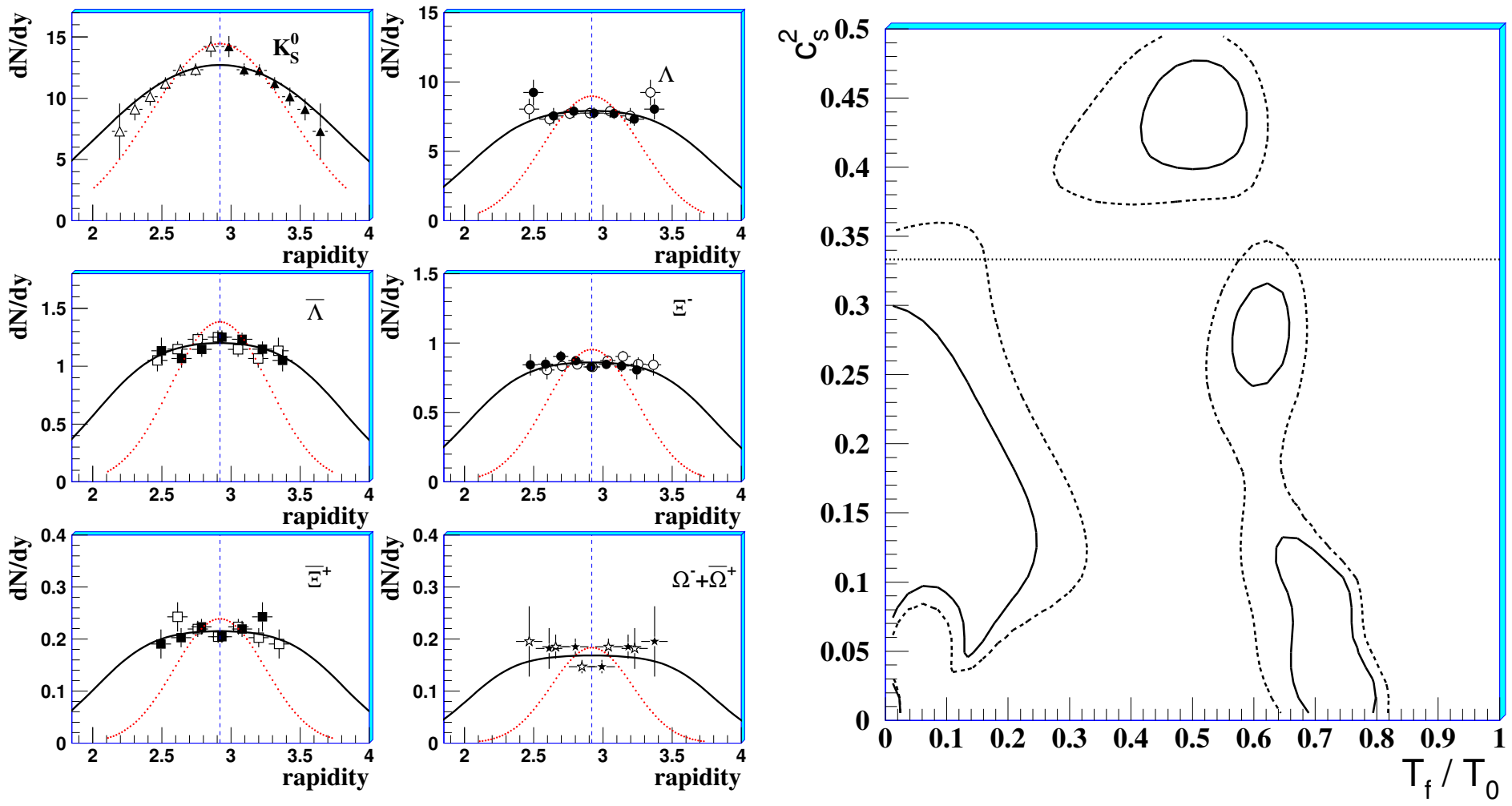

**FIGURE 3.** Left: Rapidity distributions of strange particles for the centrality range corresponding to the most central 53% of the inelastic Pb-Pb cross-section as compared to the thermal model calculation (dotted lines, in red) and a thermal model with Bjorken longitudinal flow (full lines, in black). Right: The square of the speed of sound in the medium (in unit of $c^2$) versus the ratio of the freeze-out temperature to the inital temperature. The $1\sigma$ (full curves) and the $3\sigma$ (dashed curves) confidence contours are shown. The dotted line at $c_s^2 = 1/3$ show the ideal gas limit.

## $R_{\rm CP}$ NUCLEAR MODIFICATION FACTORS

The main results on the nuclear modification factors at 160 $A$ GeV/$c$ (i.e. $\sqrt{s_{NN}}$=17.3 GeV) are discussed in this section. The complete sets of results and the details of the analysis can be found in ref. [14].

The central-to-peripheral nuclear modification factor is defined as

$$R_{\rm CP}(p_{\rm T}) = \frac{\langle N_{\rm coll}\rangle_{\rm P}}{\langle N_{\rm coll}\rangle_{\rm C}} \times \frac{{\rm d}^2N^{\rm C}_{\rm AA}/{\rm d}p_{\rm T}{\rm d}y}{{\rm d}^2N^{\rm P}_{\rm AA}/{\rm d}p_{\rm T}{\rm d}y}, \tag{2}$$

where $\langle N_{\rm coll}\rangle_{\rm C}$ and $\langle N_{\rm coll}\rangle_{\rm P}$ are the average numbers of nucleon–nucleon (NN) collisions for *central* ($C$) and *peripheral* ($P$) classes of collisions. The factor would be equal to unity if the Pb-Pb collision were a mere superposition of $N_{\rm coll}$ independent nucleon–nucleon collisions. The collision centrality is determined using the charged particle multiplicity $N_{\rm ch}$ in the pseudorapidity range $2 < \eta < 4$, sampled by the microstrip silicon detectors (MSD) as described in [15]. In table 1 we define the centrality classes used in this analysis along with the corresponding values of $\langle N_{\rm part}\rangle$ and $\langle N_{\rm coll}\rangle$ with their systematic errors. We use class 40–55% as the reference peripheral class in the denominator of $R_{\rm CP}$, see Eq. (2), and vary the 'central' class in the numerator from 0–5% to 30–40%; the results are shown in fig. 4. The shaded bands centered at $R_{\rm CP} = 1$ represent the $p_{\rm T}$-independent systematic error due to the uncertainty in the ratio $\langle N_{\rm coll}\rangle_{\rm P} / \langle N_{\rm coll}\rangle_{\rm C}$, while the shaded bands at low $p_{\rm T}$ represent the $R_{\rm CP}$ values corresponding to $N_{\rm part}$-scaling, with the band indicating the systematic error due to the uncertainty in the ratio $\langle N_{\rm part}\rangle_{\rm C} / \langle N_{\rm part}\rangle_{\rm P}$. In fig. 5 we compare our results to $R_{\rm CP}$ measurements at the SPS and at RHIC. In the left-hand panel, the WA98 $\pi^0$ data [16] for the ratio 1–6%/22–43% in Pb–Pb collisions at $\sqrt{s_{\rm NN}} = 17.3$ GeV are plotted together with the NA57 $h^-$ and $\rm K^0_S$ data for the same centrality classes. The $\rm K^0_S$ $R_{\rm CP}$ is approximately constant at 0.9 for $p_{\rm T} > 1$ GeV$/c$ and is significantly larger than that measured by the WA98 Collaboration for $\pi^0$ ($R_{\rm CP} \approx 0.6$), even when taking into account the normalization systematic errors, independent for the two experiments. The $h^-$ data from NA57 are compatible, within the systematic errors, with the $\pi^0$ data from WA98 for $p_T \lesssim 1.5$ GeV$/c$, where the $h^-$ sample is expected to be dominated by $\pi^-$. For higher $p_T$, $h^-$ have a larger $R_{CP}$ than $\pi^0$; this may be due to increasing contributions from $K^-$ and $\overline{p}$ in the $h^-$ sample. The comparison for $K^0_S$ and $\Lambda$ at SPS and RHIC (STAR data for Au–Au at $\sqrt{s_{NN}} = 200$ GeV [17]) is presented in the right-hand panel of Fig. 4. In the $p_T$ range covered by our data, up to 4 GeV$/c$, the *relative* pattern for $K^0_S$ and $\Lambda$ is similar at the two energies, while absolute values are higher at SPS than at RHIC. At RHIC, the larger $R_{CP}$ for $\Lambda$ with respect to kaons [17] or, more generally, for baryons with respect to mesons [18], in the intermediate $p_T$ range, 2–4 GeV$/c$, has been interpreted as due to parton coalescence in a high-density medium with partonic degrees of freedom. Our data show that a similar $\Lambda$–K pattern is present also at $\sqrt{s_{NN}} = 17.3$ GeV. We note that in our case such a pattern may also be explained in terms of larger Cronin effect for $\Lambda$ with respect to kaons. We have compared our results with perturbative-QCD-based theoretichal predictions [19, 20] with and without in-medium energy loss, which include the initial-state partonic intrinsic $p_T$ broadening tuned on the original Cronin effect data [21]. The calculations with in-medium energy loss describe the data much better [14].

**TABLE 1.** Average number of participants and of NN collisions with their systematic errors.

| Class (% $\sigma_{\rm inel}^{\rm Pb-Pb}$) | 0–5.0% | 10.0–20.0% | 20.0–30.0% | 30.0–40.0% | 40.0–55.0% |
|---|---|---|---|---|---|
| , $\langle N_{\rm part}\rangle$ | $345.3 \pm 1.7$ | $214.7 \pm 5.8$ | $143.0 \pm 6.6$ | $92.6 \pm 6.4$ | $49.5 \pm 5.0$ |
| $\langle N_{\rm coll}\rangle$ | $779.2 \pm 26.6$ | $421.7 \pm 26.1$ | $247.7 \pm 21.5$ | $140.5 \pm 16.2$ | $63.8 \pm 9.8$ |

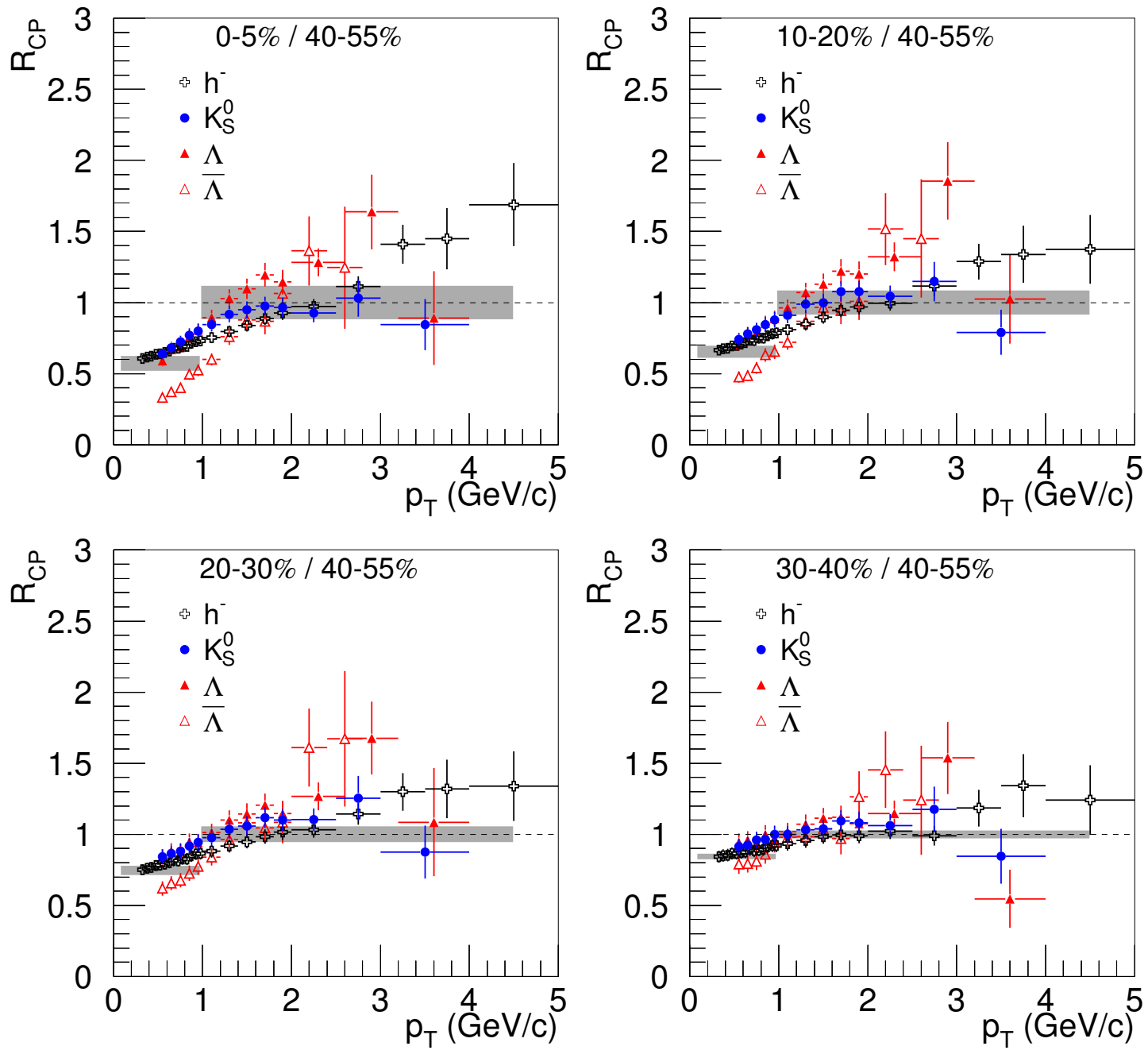

**FIGURE 4.** (colour online) Centrality dependence of $R_{\rm CP}(p_{\rm T})$ for $h^-$, $\rm K^0_S$, $\Lambda$ and $\overline{\Lambda}$ in Pb–Pb collisions at $\sqrt{s_{\rm NN}} = 17.3$ GeV. Shaded bands centered at $R_{\rm CP} = 1$ represent the systematic error due to the uncertainty in the ratio of the values of $\langle N_{\rm coll}\rangle$ in each class; shaded bands at low $p_{\rm T}$ represent the values expected for scaling with the number of participants, together with their systematic error.

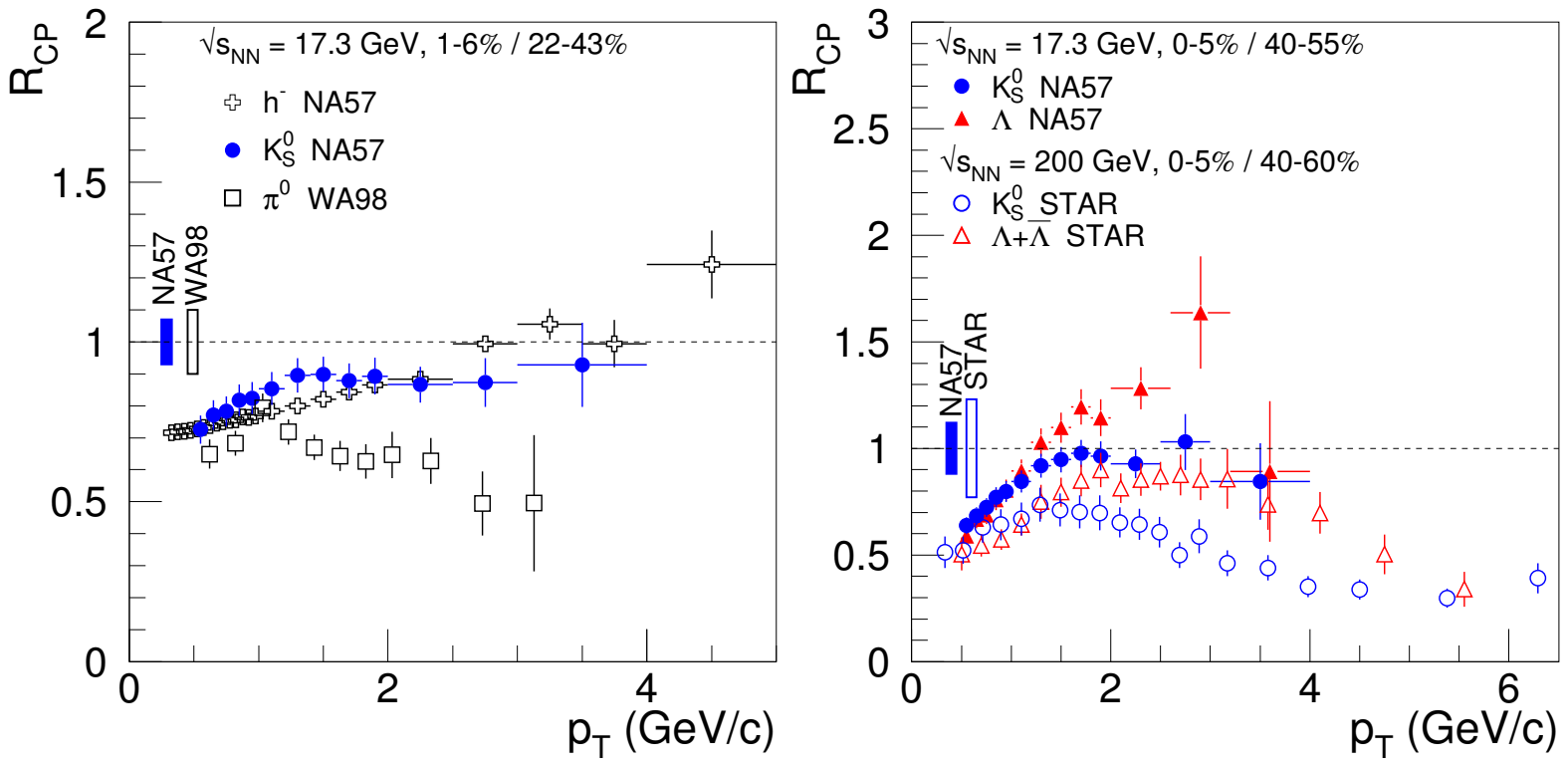

**FIGURE 5.** (colour online) Left: $R_{CP}(p_T)$ for $h^-$ and $K^0_S$ from NA57 and $\pi^0$ from WA98 [16] in Pb–Pb collisions at $\sqrt{s_{NN}} = 17.3$ GeV. Right: $R_{CP}(p_T)$ for $K^0_S$ and $\Lambda$ in Pb–Pb at $\sqrt{s_{NN}} = 17.3$ GeV (NA57) and in Au–Au at $\sqrt{s_{NN}} = 200$ GeV (STAR) [17]; slightly different peripheral classes are employed for the comparison. The bars centered at $R_{CP} = 1$ represent the normalization errors; the point-by-point bars are the quadratic sum of statistical and systematic errors.

# CONCLUSION

We have measured the $dN/dp_T$ and $dN/dy$ distributions of high purity samples of $K^0_S$, $\Lambda$, $\Xi$ and $\Omega$ particles produced at central rapidity in Pb-Pb collisions over a wide centrality range of collision (i.e. the most central 53% of the Pb–Pb inelastic cross-section).

The analysis of the transverse mass spectra of strange particles in Pb-Pb collisions at SPS energies suggests that after a central collision the system expands explosively and then it freezes-out when the temperature is of the order of 120 MeV, with an average transverse flow velocity of about one half of the speed of light. Similar transverse flow velocities are measured at 40 and 158 $A$ GeV/$c$ but the freeze-out temperature is lower at the lower energy. The results on the centrality dependence of the expansion dynamics indicate that with increasing centrality the transverse flow velocity increases and the freeze-out temperature decreases.

Boost-invariant Bjorken hydrodynamics can describe simultaneously the rapidity spectra of all the strange particles under study with $\chi^2/ndf \approx 1$, yielding an average longitudinal flow velocity $< \beta_L >= 0.42 \pm 0.03$, sligthly larger than the measured transverse flow. The *longitudinal* flow velocity being close to the transverse one is suggestive of large nuclear stopping. A fairly good description is also provided by Landau hydrodynamics, which allows us to put constraints in the parameter space of the speed of sound in the medium and the ratio of the freeze-out temperature to the initial temperature.

Central-to-peripheral nuclear modification factors for $K^0_S$, $\Lambda$, $\overline{\Lambda}$ and $h^-$ in Pb–Pb collisions at top SPS energy have been measured as a function of $p_T$ up to about 4 GeV/$c$. At low $p_T$, $R_{CP}$ agrees with $N_{part}$ scaling for all the particles under consideration, except the $\overline{\Lambda}$, for which the yields at low $p_T$ are found to increase slower than the number of participants. For $p_T \gtrsim 1$ GeV/$c$, $K^0_S$, $\Lambda$ and $\overline{\Lambda}$ show a pattern similar to that observed in Au–Au collisions at top RHIC energy, although the $R_{CP}$ values are found to be larger at SPS. At RHIC, this pattern has been interpreted in the framework of models that combine parton energy loss with hadronization via coalescence, at intermediate $p_T$, and via fragmentation, at higher $p_T$. The measured $K^0_S$ 0–5%/40–55% $R_{CP}$ is not reproduced by a theoretical calculation that includes only initial-state nuclear effects. The data can be better described by including final-state parton energy loss as predicted for SPS energy on the basis of RHIC data.